\documentclass[12pt,twocolumn]{article}
\usepackage{graphicx}
\usepackage{color}
\usepackage{hyperref}
\usepackage{caption}
\usepackage{setspace}
\usepackage[margin=1in]{geometry}
\usepackage{ulem}
\usepackage[T1]{fontenc}
\usepackage{amsmath}
\usepackage[utf8]{inputenc}
\usepackage{hyperref}
\usepackage{epstopdf}
\onecolumn
\begin{document}
\begin{center}
{\Large {\bf Effect of Random-Bond Disorder on KT Transition }}
\end{center}
\vskip 0.5cm
\begin{center}
{\it Nepal Banerjee} \\
{\it Department of Physics,University of Seoul,South Korea}
\vskip 0.2 cm
{\bf {Email}}: nb.uos1989@gmail.com
\end{center}
\vskip 0.5cm
\begin{abstract}
Here we have simulated effect of quenched type random-bond disorder during the XY transition.Here we have studied the spontaneous magnetization(M),heat-capacity(Cv) with T.Here we notice a spontaneous symmetry breaking and  observe quasi long range order (QLRO) at ground state in presence of this type of bond-random disorder.

\end{abstract}
\section{Introduction}
Several breakthrough discovery on 2D van der Waals(vdW) magnet introduced to us a challenging and very complex phase of matter\cite{ajayan2016van,jung2015origin,song,blanet}.High tunibality of this materials able to probe several exotic correlated phase\cite{chen2019evidence,jung2014ab,burch2018magnetism,nitin, gong2017discovery}.Kosterlitz-Thouless(KT) transition in this van der Waals magnetic material turn into a very exciting research field from last few years\cite{kosterlitz1, kosterlitz2,MA,Hasen,kosterlitz2018topological,  Li,s_das}.Developing the quasi long range order (QLRO) during the transition  still remain a mysterious phenomena for this complex material.Several vdW materials hosting this type of transition and we know that real materials are full of imperfection and intrinsic disorder and that made this types of transition a most striking phenomena\cite{chit2,cri3_1,KB}.
\begin{figure}[htp]
\centering
\includegraphics[scale=0.28]{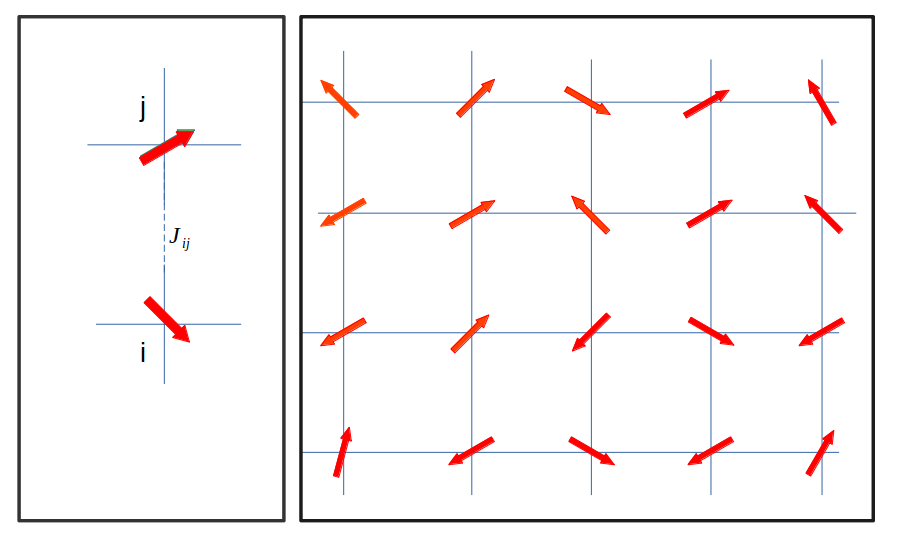}
\caption{A Schematic picture of XY spins in square lattice.Here each spin interacting with its neighbour with random exchange interaction $J_{ij}$.This random exchange interaction  $J_{ij}$ vary randomly site to site and it lies in a particular range between 0 to 2 and it can take any random value uniformly between this range.}
\label{}
\end{figure}
Here we have been trying to simulate  XY transition in presence of simulated disorder which is a random-bond type quenched disorder\cite{
Haris,Imry,vojta2019disorder,okabe}.Here we have planned our paper in a following way.Firstly we have presented model Hamiltonian and then presented the simulation methodology and results.Finally we have made conclusion after discussing the results.

\section{Model Hamiltonian}
Here we are describing our model Hamiltonian,which is a random-bond disorder induced 2D XY model in a square lattice\cite{tobochnik1979monte}.Here we consider that the exchange interaction $J(i \rightarrow j) \neq J(j \rightarrow i)$,where i and j indicating two different nearest-neighbour lattice site.
\begin{eqnarray}
H=-\sum_{<i,j>}  J_{ij}^x  S_i^x  S_j^x + J_{ij}^y  S_i^y  S_j^y 
\end{eqnarray}
Here $\vec S_i=(S_i^x,S_i^y)$ represent classical two component spin with unit magnitude sitting at each lattice point i and $J_{ij}^x$ and $J_{ij}^y$ are random exchange interaction through which simulated XY spins are interacting with its nearest-neighbour spins(Fig:1).Here we are introducing a spin model where each spin interacting with its neighbour spins with a specific random exchange interaction and that interaction strength vary randomly between a particular range from 0 to 2 uniformly.

\section{Simulation Methodology and Results}
Here we are using the classical Monte-Carlo simulation technique for simulation of disorder induced 2D XY magnet.Here we have simulated the XY type of spin at each site of 2D lattice grid which have two component and we have used the following formula for this simulation\cite{drouin2022kosterlitz}.

\begin{eqnarray}
S_x &=&|S|\cos(\phi)\\
S_y &=&|S|\sin(\phi) 
\end{eqnarray}
Here we have chosen the  magnitude of spin $|S|=1$ and we have simulated the different component of spin after choosing the $\phi$  randomly where $\phi $ vary from 0 to $2\pi$. 
\begin{figure}[htp]
\centering
\includegraphics[scale=0.30]{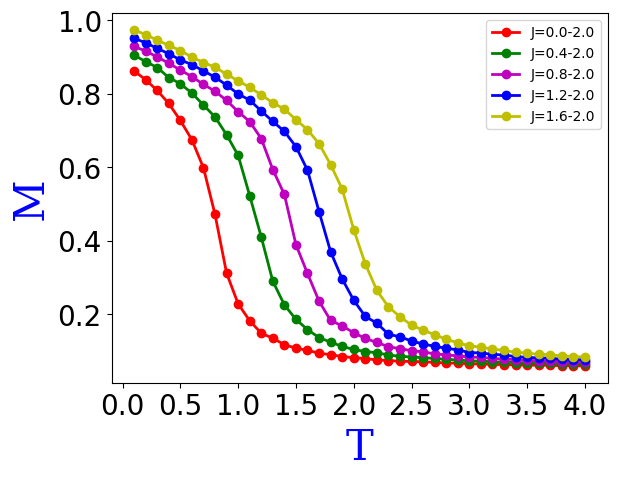}
\includegraphics[scale=0.30]{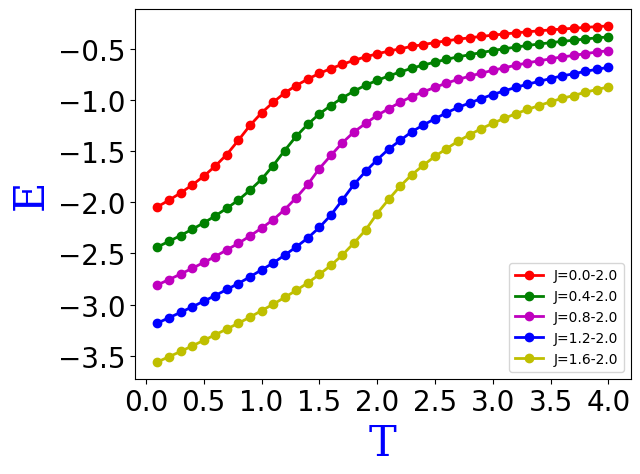}
\includegraphics[scale=0.30]{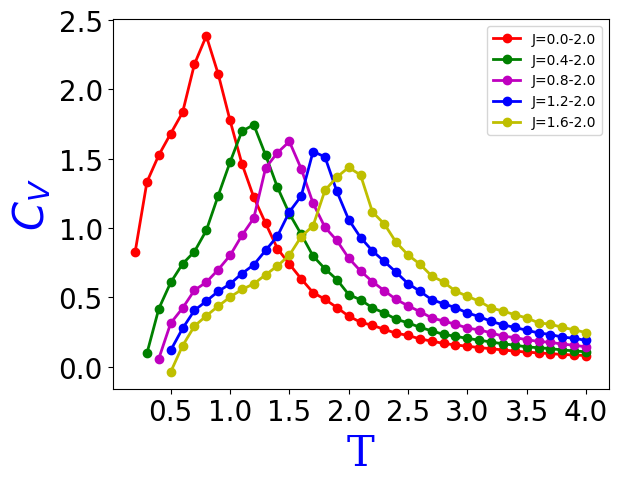}
\caption{Here we have presented the spontaneous magnetization(M) and heat-capacity(Cv) with T.Here we have presented the results for 
$20\times 20$ lattice size at different strength of random-bond disorder.Here for simplicity we consider Boltzmann constant $k_B=1$.Here $J_{ij}^x,J_{ij}^y$ vary randomly from 0.0-2.0, 0.4-2.0 ,0.8-2.0, 1.2-2.0 ,1.6-2.0.Here magnitude of spin is |S|=1.}
\label{}
\end{figure}
Here we have introduced the random-bond type disorder and simultaneously simulating the exchange-interaction through which simulated XY spin are interacting with its nearest-neighbour spin.In our simulation we consider $J(i \rightarrow j ) \neq J( j \rightarrow i )$.Here i and j are two nearest neighbour.In this simulation we have used single spin flipping Metropolis algorithm and we have used $10^5$ MC steps for creating equivalent ensembles and among them we only consider $5 \times 10 ^4$ ensembles for taking the average of different thermodynamic quantity.Here we have calculated different thermodynamic quantity like spontaneous magnetization(M),heat-capacity($C_V$) and energy(E) at different T.We have used the following formula for the calculation of those thermodynamic quantity.
\begin{eqnarray}
M &=&\sqrt{m_x^2 +m_y^2 }\\
{C_V}&=& L^2(<E^2> - <E>^2 )
\end{eqnarray} 
Here $m_x=\sum S_x/L^2$,$m_y=\sum S_y/L^2$.Here we have used $20 \times 20 $ lattice grid for this simulation.Here we have noticed interesting behaviour of M at different range of exchange interaction strength which is lies between 0.0-2.0, 0.4-2.0, 0.8-2.0, 1.2-2.0, 1.6-2.0.Here we have notice a significant change of scaling behaviour of M at different range of exchange interaction.Here we are noticing that the transition temperature is decreasing because of bond dilution and that is revealing in the behaviour of M  and $C_V$ with T.Here we also notice the significant change of  heat-capacity($C_V$) with T and we notice that the scaling behaviour of $C_V$ is changing with T.We have noticed that the peak of $ C_V $ is showing more sharp peak when we increasing the range of random-bond disorder and Tc is shifting towards lower temperature.Here we are observing ferromagnetic type QLRO as a ground state for all the realization of random-bond disorder\cite{banerjee2023simulation,olivia}.
\section{Discussion and conclusion}
Here we have presented the results of random-bond type quenched disorder in a XY model.We have presented the temperature evolution spontaneous magnetization (M) and heat-capacity($C_V$) with T.We observed that because of dilution of bond the Tc is not only reduced but the scaling behaviour of that phase transition is also changed in presence of this type of disorder.
\section{Acknowledgement}
We are greatly acknowledging IIT kanpur for giving visiting scholar position and providing generous research facility,financial support and hospitality during the visit.Author is greatly acknowledging Prof.Jeil Jung and Prof.M.Acharyya for several helpful guidance during the progress of this project.


\begin{thebibliography}{}

\bibitem{ajayan2016van}Pulickel Ajayan, Philip Kim, and Kaustav Banerjee. van der waals materials.Phys. Today, 69(9):38, 2016.

\bibitem{jung2015origin} Jeil Jung, Ashley M DaSilva, Allan H MacDonald, and Shaffique Adam. Origin of band gaps in graphene on hexagonal boron nitride.Nature communications, 6(1):6308, 2015. 

\bibitem{song}Song, Tiancheng, et al. Direct visualization of magnetic domains and moiré magnetism in twisted 2D magnets. Science 374.6571 (2021): 1140-1144.

\bibitem{blanet}Hejazi, Kasra, Zhu-Xi Luo, and Leon Balents. Noncollinear phases in moiré magnets. Proceedings of the National Academy of Sciences 117.20 (2020): 10721-10726.
 
\bibitem{chen2019evidence}Guorui Chen, Lili Jiang, Shuang Wu, Bosai Lyu, Hongyuan Li,Bheema Lingam Chittari,Kenji Watanabe, Takashi Taniguchi, Zhiwen
Shi, Jeil Jung, et al. Evidence of a gate-tunable mott insulator in a trilayer graphene moiré superlattice.Nature Physics, 15(3):237-241, 2019

\bibitem{jung2014ab}Jeil Jung, Arnaud Raoux, Zhenhua Qiao, and Allan H MacDonald. Ab initio theory of moiré superlattice bands in layered two-dimensional materials. Physical Review B, 89(20):205414,2014.

\bibitem{burch2018magnetism}Kenneth S Burch, David Mandrus, and Je-Geun Park. Magnetism in two-dimensional
van der waals materials. Nature, 563(7729):47-52, 2018.

\bibitem{nitin}Samarth, N. (2017). Magnetism in flatland. Nature, 546(7657), 216-217.

\bibitem{gong2017discovery}Cheng Gong, Lin Li, Zhenglu Li, Huiwen Ji, Alex Stern, Yang Xia, Ting Cao, Wei
Bao, Chenzhe Wang, Yuan Wang, et al. Discovery of intrinsic ferromagnetism in two-
dimensional van der waals crystals. Nature, 546(7657):265-269, 2017.

\bibitem{kosterlitz1}John Michael Kosterlitz and David James Thouless. Ordering, metastability and phase
transitions in two-dimensional systems. Journal of Physics C: Solid State Physics,
6(7):1181, 1973.

\bibitem{kosterlitz2}Enzo Granato and JM Kosterlitz. Critical behavior of coupled xy models. Physical
Review B, 33(7):4767, 1986.

\bibitem{MA}Muktish Acharyya and Erol Vatansever. Monte carlo study of the phase diagram of
layered xy antiferromagnet. Physica A: Statistical Mechanics and its Applications,
605:128018, 2022.

\bibitem{Hasen}Martin Hasenbusch. A monte carlo study of the three-dimensional xy universality class:
Universal amplitude ratios. Journal of Statistical Mechanics: Theory and Experiment,
2008(12):P12006, 2008.

\bibitem{kosterlitz2018topological} John Michael Kosterlitz. Topological defects and phase transitions.
Journal of Modern Physics B, 32(13), 2018.


\bibitem{Li}Li, H., Liao, Y. D., Chen, B. B., Zeng, X. T., Sheng, X. L., Qi, Y., ...  Li, W. (2020). Kosterlitz-Thouless melting of magnetic order in the triangular quantum Ising material TmMgGaO4. Nature communications, 11(1), 1111.

\bibitem{s_das}Das, S., Voleti, S., Saha-Dasgupta, T.,  Paramekanti, A. XY magnetism, Kitaev exchange, and long-range frustration in the J eff= 1 2 honeycomb cobaltates. Physical Review B, 104(13), 134425,2021.


\bibitem{chit2}Bheema Lingam Chittari,Dongkyu Lee,Nepal Banerjee, Allan H MacDonald, Euyheon Hwang, and Jeil Jung. Carrier-and strain-tunable intrinsic magnetism in two-dimensional $MAX_3$ transition metal chalcogenides. Physical Review B, 101(8):085415,2020.

\bibitem{cri3_1}Chao Lei, Bheema L Chittari, Kentaro Nomura, Nepal Banerjee, Jeil Jung, and Allan H MacDonald. Magnetoelectric response of antiferromagnetic $CrI_3$ bilayers.Nano Letters,21(5):1948-1954, 2021.

\bibitem{KB}KB Yogendra, Tanmoy Das, and G Baskaran. Emergent glassiness in disorder-free kitaev model. arXiv preprint arXiv:2302.14328, 2023.


\bibitem{Haris}Harris, A. B.Effect of random defects on the critical behaviour of Ising models. Journal of Physics C: Solid State Physics, 7(9), 1671,1974

\bibitem{Imry}Imry, Y.,  Ma, S. K. Random-field instability of the ordered state of continuous symmetry. Physical Review Letters, 35(21), 1399, 1975

\bibitem{vojta2019disorder}Thomas Vojta. Disorder in quantum many-body systems. Annual Review of Condensed Matter Physics, 10:233-252, 2019.

\bibitem{okabe}Surungan, T.,Okabe, Y. (2005). Kosterlitz-Thouless transition in planar spin models with bond dilution. Physical Review B, 71(18), 184438.


\bibitem{tobochnik1979monte}Jan Tobochnik and GV Chester. Monte carlo study of the planar spin model.
Physical Review B, 20(9):3761, 1979.

\bibitem{drouin2022kosterlitz}Victor Drouin-Touchette. The kosterlitz-thouless phase transition: an introduction for the intrepid student. arXiv preprint arXiv:2207.13748, 2022.


\bibitem{banerjee2023simulation}Nepal Banerjee. Simulation of kosterlitz-thouless (KT) transition with classical monte-
carlo simulation. arXiv preprint arXiv:2307.10310, 2023.


\bibitem{olivia}Mallick, O.,Acharyya,M. Monte Carlo study of the phase transitions in the classical XY ferromagnets with random anisotropy. Phase Transitions, 1-19,2023.

\end{thebibliography}
\end{document}